\documentclass[prl,twocolumn,showpacs,floatfix]{revtex4}
\usepackage{times}

\usepackage{graphicx}
  \graphicspath{{./Eps/}}
\usepackage{amsmath}
\usepackage{amssymb}

\newcommand{\vc}[1]{
\textbf{\textit{#1}} }

\begin{document}

\title{Coexistence of diffusive resistance and ballistic persistent current in disordered metallic rings with rough edges: Possible origin of puzzling experimental values}

\author{J. \surname{Feilhauer}}

\author{M. \surname{Mo\v{s}ko}}
\email{martin.mosko@savba.sk}

\affiliation{Institute of Electrical Engineering, Slovak
Academy of Sciences, 841 04 Bratislava, Slovakia}

\date{\today}

\begin{abstract}
Typical persistent current ($I_{typ}$) in a mesoscopic normal metal ring with disorder due to rough edges and random grain boundaries is calculated by a scattering matrix method. In addition, resistance of a corresponding metallic wire is obtained from the Landauer formula and the electron mean free path ($l$) is determined. If disorder is due to the rough edges, a ballistic persistent current $I_{typ} \simeq e v_F/L$ is found to coexist with the diffusive resistance ($ \propto L/l$), where $v_F$
is the Fermi velocity and $L \gg l$ is the ring length. This ballistic current is due to a single electron that moves almost in parallel with the rough edges and thus hits them rarely (it is shown that this parallel motion exists in the ring geometry owing to the Hartree-Fock interaction). Our finding agrees with
a puzzling experimental result $I_{typ}\simeq e v_F/L$, reported by Chandrasekhar et al. [Phys. Rev. Lett. \textbf{67}, 3578 (1991)] for metallic rings of length $L \simeq 100 l$. If disorder is due to the grain boundaries, our data reproduce theoretical result $I_{typ}\simeq (e v_F/L) (l/L)$ that holds for the white-noise-like disorder and has been observed in recent experiments. Thus,
result $I_{typ}\simeq e v_F/L$ in a disordered metallic ring of length $L \gg l$ is as normal as
result $I_{typ}\simeq (e v_F/L)(l/L)$. Which result is observed depends on the nature of disorder. Experiments that would determine
$I_{typ}$ and $l$ in correlation with the nature of disorder can be instructive.
\end{abstract}

\pacs{73.23.-b, 73.23.Ra}
\keywords{quasi one-dimensional transport, surface roughness, quantum conductance,
universal conductance fluctuations}

\maketitle


 \section{I. Introduction}

It is known that a conducting ring pierced by magnetic flux can support persistent
electron current \cite{Imry-book}. Persistent currents exist in superconducting rings \cite{Deaver}, in mesoscopic resistive metal rings \cite{Buttiker,Chand,Jariwala,Bluhm,Bles}, in ballistic metallic rings \cite{Mailly}, and in nanorings made of band insulators \cite{Moskova}.

At zero temperature, the mesoscopic resistive metal ring pierced by magnetic flux $\Phi$ supports the persistent current
$I = {\sum}_{ \forall E_j \leq E_F} I_j$, where $I_j(\Phi) = -dE_j(\Phi)/d\Phi$ is the current carried by the electron with eigen-energy $E_j(\Phi)$, and $E_F$ is the Fermi level \cite{Buttiker,Levy,Mailly}. Function $I(\Phi)$ is periodic with
period $\Phi_0 \equiv h/e$, which is an experimental signature of the persistent current \cite{Levy,Mailly,Chand,Jariwala,Bluhm,Bles}.
If the ring is clean and possesses one conducting channel, the sum $\sum I_j$ changes its sign whenever a new occupied state $j$ is added.
Due to the sign cancelation mainly the electron at the Fermi level contributes to the sum, and the amplitude of the current is $I_0 = ev_F/L$ \cite{Cheung1D},
where $v_F$ is the Fermi velocity and $L$ the ring circumference.
If the ring is disordered, the size and sign of
the current fluctuate from sample to sample and
a typical current \emph{per} one ring is $I_{typ}=\langle I^2\rangle ^{1/2}$, where $\langle \dots \rangle$ means ensemble average.

The number of the conducting channels ($N_c$) in the disordered metallic rings is usually large ($N_c \gg 1$) and the rings obey the diffusive limit, $l \ll L \ll \xi$, where $l$ is the electron mean free path and $\xi \simeq N_c l$ is the localization length.
 To estimate $I_{typ}$, assume again that mainly the electron at the Fermi level contributes to the sum $\sum I_j$. Since $L \gg l$, the electron is expected to move around the ring by diffusion. Its transit time is $\tau_D = L^2/D$, where $D = v_F l/d$ is the diffusion coefficient and $d$ is the sample dimensionality. So $I_{typ} \simeq e/\tau_D = (1/d)(e v_F/L)(l/L)$.
  A similar result follows from the Green function theory \cite{Cheung,Riedel} which assumes the non-interacting electrons and emulates disorder by a random potential $V(\bold{r})$ obeying the white-noise condition $\langle V(\bold{r}) V(\bold{r}')\rangle \propto \delta(\bold{r} - \bold{r}')$. The theory \cite{Cheung,Riedel} gives
\begin{equation}
I^{theor}_{typ} = 2 \times (1.6/d) (e v_F/L) (l/L), \quad l \ll L \ll \xi ,
 \label{Igre}
\end{equation}
where the factor of $2$ is due to the electron spin, $d =1$, $2$, or $3$, and the origin of the factor of $1.6$ is explained in Ref. \cite{Feilhauer1}.

The first observation of the persistent current in a single metallic ring was reported \cite{Chand} for three Au rings of size $L \sim 100 l$. The
measured currents
showed the desired flux-periodicity $\Phi_0$, but they
were ten-to-hundred times larger than result \eqref{Igre}; they ranged from $\sim 0.1ev_F/L$ to $\sim ev_F/L$.
This huge discrepancy has not been explained yet \cite{Saminadayar,Shanks}.
Other Au rings showed \cite{Jariwala} the currents slightly larger than the result \eqref{Igre} and recent experiments \cite{Bluhm,Bles} confirmed the result \eqref{Igre} well.

Why did the similar measurements of diffusive Au rings \cite{Chand,Bluhm} show quite different results, $I_{typ}\simeq e v_F/L$ and
$I_{typ}\simeq (e v_F/L) (l/L)$? A puzzle \cite{Chand} is why a multichannel disordered ring of length $L \gg l$ carries the current $e v_F/L$, typical for a one-channel ballistic ring? These questions are known as unresolved problems of mesoscopic physics \cite{Saminadayar,Shanks,Bles}.

 This paper answers both questions theoretically. It is known \cite{Saminadayar} that there is disorder due to polycrystalline grains and rough edges
even in a pure Au ring. Using a single-particle scattering-matrix method \cite{Feilhauer1,Feilhauer2}, we calculate the typical persistent currents in the Au rings with grains and rough edges \emph{without the white-noise approximation}. Another key point of our single-particle approach is that \emph{our description of the single-electron states in the ring captures an essential effect of the Hartree-Fock interaction, the cancelation of the centrifugal force by an opposite oriented Hartree-Fock field}.

Our findings can be summarized as follows. If the disorder is due to the polycrystalline grains, our results agree with the white-noise-related formula \eqref{Igre} and experiments \cite{Bluhm,Bles}. However, if the disorder is due to the rough edges, we find the ballistic-like result $I_{typ}\simeq e v_F/L$ albeit the resistance is diffusive ($\propto L/l$) and $L \gg l$, like in the experiment \cite{Chand}. This ballistic current is due to a single electron that moves (almost) in parallel with the rough edges and thus hits them rarely. We show that this parallel motion exists in the ring geometry owing to the Hartree-Fock interaction. Our major message reads: result $I_{typ}\simeq e v_F/L$ in a metal ring of length $L \gg l$ is as normal as
result $I_{typ}\simeq (e v_F/L)(l/L)$. Which result is observed depends on the nature of disorder.

    We note that we focus us on the typical current rather than on the mean current $\langle I \rangle$. The sign and amplitude of the mean current measured in the experiment by Levy et al. \cite{Levy} is another puzzling problem in the field. This problem has been addressed
in reference \cite{comment1} within the interacting electron model. On the other hand, reference \cite{comment1} did not study the typical current.
 It is tempting to think that the typical current is not affected by electron-electron interaction; at least, experiments \cite{Bluhm,Bles} confirm result $I_{typ}\simeq (e v_F/L)(l/L)$ which has been derived \cite{Cheung,Riedel} for non-interacting electrons. We are thus motivated to study the typical current within a single-particle model. However, our single-particle model is not a truly non-interacting model because it captures a key
 effect of the Hartree-Fock interaction.

 Our paper is organized as follows. In section II, resistance of wires with rough edges and wires with grains is calculated by means of the scattering-matrix approach \cite{Feilhauer1,Feilhauer2,tamura,cahay,saenz}. In section III we focus us on the single-particle states in clean metal rings. We demonstrate the key role of the Hartree-Fock interaction and we provide a simple intuitive argument about the existence of ballistic
 current $I_{typ}\simeq e v_F/L$ in rings with rough edges. Microscopic calculations of persistent currents are presented in section IV. Finally, in section V a summary of our work is given with a few concluding remarks.

\section{II. Resistance of wires with grain boundaries and wires with rough edges}

For simplicity, we study two-dimensional (2D) rings and discuss the 3D effects briefly at the end of the paper. Experimentally \cite{Levy,Mailly,Chand,Jariwala,Bluhm,Bles}, persistent currents in rings were studied together with the resistance of the co-deposited wires in order to determine the mean free path $l$. In this section we study the wire resistance and mean free path. Sections II.A and II.B describe our transport model and our results, respectively. Of special importance is section II.C. It shows that our edge-roughness model gives the transport results which are universal - independent on the choice of the roughness model.

\subsection{A. Transport model}

\begin{figure}[t!]
\centerline{\includegraphics[clip,width=0.98\linewidth]{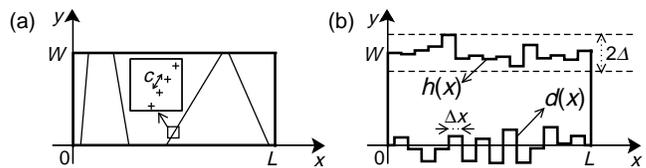}}
\caption{Our models of disordered wires: (a) wire with grain boundaries, (b) wire with rough edges. The meaning of all symbols used in the figure is described in the main text.} \label{Fig-1}
\end{figure}

We consider a stripe-shaped 2D wire (Fig. \ref{Fig-1}) described by
Hamiltonian \cite{Feilhauer1,Feilhauer2}
\begin{equation}
 H = - \frac{\hbar^2}{2m^*} \left( \frac{\partial^2}{\partial x^2} + \frac{\partial^2}{\partial y^2}\right) + U \left(x,y \right)
+ V \left( x,y \right) ,
\label{hamiltdisord}
\end{equation}
where $m^*$ is the electron effective mass, $U$ is the grain boundary potential, and $V$ is the potential due to the wire edges.
To simulate the electron transport in wires with grain boundaries, we will rely on the scattering matrix approach developed in the works \cite{Feilhauer2,tamura,cahay}. Similarly, to simulate the electron transport in wires with rough edges, we will rely on the scattering matrix approach described in the works \cite{Feilhauer1,saenz}. Here we review both approaches briefly by means of figure \ref{Fig-1}.

Let $d(x)$ and $h(x)$ be the $y$-coordinates of the edges. Then
 \begin{eqnarray}
V(x,y) = \left\{
                          \begin{array}{ll}
                            0, & d(x)<y<h(x) \\
                            \infty, & \mbox{elsewhere}
                          \end{array} \right.
.
\label{smoothpot1}
\end{eqnarray}
For smooth edges one has $d(x)=0$ and $h(x)=W$, while in the case of the rough edges $d(x)$ and $h(x)$ fluctuate randomly in the intervals $\langle -\Delta, \Delta \rangle$ and $\langle W-\Delta, W+\Delta \rangle$, respectively. It can be shown \cite{Feilhauer1} that the RMS of such random fluctuations ($\delta$) is simply $\delta = \Delta/\sqrt{3}$. The fluctuations are assumed to appear along the edges abruptly with a constant step $\Delta x$ which plays (within this model) the role of the roughness correlation length \cite{Feilhauer1}. The parameters of our roughness model are thus $\delta$ and $\Delta x$.
 The grain boundaries are modeled as a randomly-oriented mutually non-intersecting lines, where the angle between the line and $x$-axis is random \cite{Feilhauer2}. Each line  consists of equidistant repulsive dots (depicted by the plus signs) with potentials $\gamma \delta(x-x_i) \delta(y - y_i)$, where ($x_i$, $y_i$) is the position of the $i$-th dot. Thus
$U(x,y) = {\sum}_{ \forall i} \gamma \delta(x - x_i) \delta(y - y_i)$. If the inter-dot distance $c$ approaches zero and the ratio $\gamma/c$ is fixed, a grain boundary scatters electrons as a structure-less
 line-shaped barrier independent on the choice of $c$. If a 2D electron impinges on such a barrier perpendicularly with Fermi wave vector $k_F$, it is reflected with probability \cite{Feilhauer2}
\begin{equation}
R_G = (\bar{\gamma}/c)^2/[k^2_F + (\bar{\gamma}/c)^2],
\label{reflexboundary}
\end{equation}
where $\bar{\gamma} = m^* \gamma / \hbar^2$. The parameters of our grain boundary model are the reflection probability $R_G$ (typically \cite{comment00} $R_G \sim 0.1 - 0.8$) and the mean inter-boundary distance $d_G$.

We connect the wire to two ideal leads - clean long wires of width $W$.
The spectrum of the electron wave functions $\psi (x,y)$ and electron energies $E$ in the leads is given by
\begin{equation} \label{subpasovafunkcia}
\psi (x,y) = \\ e^{ikx} \chi_n(y), \quad n = 1,2, \dots \infty,
\end{equation}
and
\begin{equation}
E = \epsilon_n + \frac{\hbar^2}{2m^*}k^2, \quad \epsilon_n \equiv \frac{\hbar^2 \pi^2}{2m^*W^2} n^2,
\label{spasy}
\end{equation}
where $k$ is the electron wave vector in the $x$ direction, $\epsilon_n$
is the eigen-energy of motion in the $y$-direction, and
\begin{equation}
\begin{array}{c}
\chi_n(y) = \left\{
                          \begin{array}{ll}
                            \sqrt{\frac{2}{W}} \sin \left( \frac{\pi n}{W}y \right), & 0<y<W \\
                            0, & \mbox{elsewhere}
                          \end{array} \right.
\end{array}
\label{subbandfunction}
\end{equation}
is the wave function in direction $y$. Thus, in the leads we have for the electron energy $E$ a general wave function \cite{Datta-kniha,Feilhauer1,Feilhauer2}
\begin{equation}
\begin{array}{c}
\psi(x,y) = {\sum}^{{}_N}_{n=1} \left[ A^+_n(x) + A^-_n(x) \right] \sin(\frac{n \pi y}{W}), \ x\leq 0 \\
\psi(x,y) = {\sum}^{{}_N}_{n=1} \left[ B^+_n(x) + B^-_n(x) \right] \sin(\frac{n \pi y}{W}), \ x \geq L
\end{array}
\label{rozvoje}
\end{equation}
 where $N$ is the considered number of channels (ideally $N = \infty$), $A^{\pm}_n(x) \equiv a^{\pm}_{n} e^{\pm i k_n x}$, $B^{\pm}_n(x) \equiv b^{\pm}_{n} e^{{\pm} i k_n x}$, and $k_n(E)$ is the wave vector given by equation $\frac{\hbar^2 k^2}{2m^*}+\frac{\hbar^2 \pi^2 n^2}{2m^*W^2}=E$.
Vectors $\vc{A}^\pm(0)$ and $\vc{B}^\pm(L)$ with components $A^\pm_{n=1, \dots N}(0)$ and $B^\pm_{n=1, \dots N}(L)$
obey the matrix equation \cite{Datta-kniha,Feilhauer1,Feilhauer2,tamura,cahay,saenz}
  %
\begin{eqnarray}
\left(
\begin{array}{c}
\vc{A}^-(0) \\
\vc{B}^+(L) \\
\end{array}
\right)
=
\left[
\begin{array}{cc}
r & t' \\
t & r' \\
\end{array}
\right]
\left(
\begin{array}{c}
\vc{A}^+(0) \\
\vc{B}^-(L) \\
\end{array}
\right),
\quad
S
\equiv
\left[
\begin{array}{cc}
r & t' \\
t & r' \\
\end{array}
\right] ,
\label{Smatrixrovnica}
\end{eqnarray}
where $S$ is the scattering matrix \cite{Datta-kniha}. Its elements $t(E)$, $r(E)$, $t'(E)$, and $r'(E)$ are matrices with dimensions $N \times N$. Matrices $t$ and $t'$ are the transmission amplitudes of the waves $\vc{A}^+$ and $\vc{B}^-$, respectively, and matrices $r$ and $r'$ are the corresponding reflection amplitudes. In particular, the matrix element $t_{mn}(E)$ is the transmission amplitude from channel $n$ in the left lead into the channel $m$ in the right lead. We evaluate $S(E)$ for disorder in figure \ref{Fig-1} by methods of papers \cite{Feilhauer1,Feilhauer2}.

At zero temperature, the wire conductance $g$ (in units $2e^2/h$) is given by the Landauer formula
$g = \sum^{N_c}_{n=1} T_{n}$,
where
\begin{equation}
T_{n}(E_F)  = \sum^{N_c}_{m=1} |t_{mn}(E_F)|^2 \frac{k_m(E_F)}{k_n(E_F)}
\label{transmisia}
\end{equation}
is the transmission probability of channel $n$.
 We evaluate $t_{mn}$ for a large statistical ensemble of samples \cite{Feilhauer1,Feilhauer2} and obtain the mean transmission $\langle T_n \rangle$ and mean resistance $\langle \rho \rangle = \langle 1/g \rangle$.

\subsection{B. Transport results}

\begin{figure}[t!]
\centerline{\includegraphics[clip,width=0.98\columnwidth]{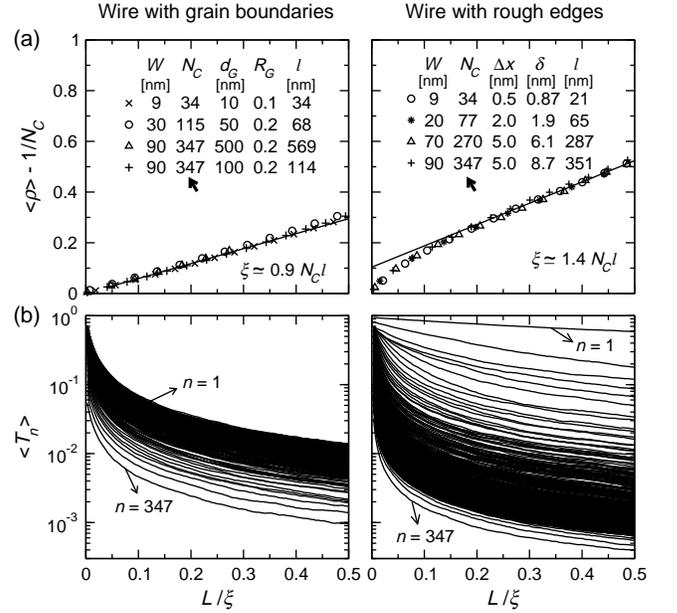}}
\vspace{-0.15cm} \caption{Transport in disordered Au wires. Parameters of Au are $m^* = 9.1 \times 10^{-31}$kg and $E_F = 5.6$eV, other parameters are listed. Not to affect the results, in our calculations $N$ is usually kept larger than $N_c$.
Figure (a) shows the mean resistance $\langle \rho \rangle$ versus $L$.
Note that $\langle \rho \rangle$ is reduced by resistance $1/N_c$ and $L$ scaled by $\xi$. The localization length $\xi$ is obtained
\cite{Feilhauer1,Feilhauer2} from numerical data for $\langle \ln g \rangle$ by using the fit $\langle \ln g \rangle = - L/\xi$ at $L \gg \xi$. The full lines show the linear fit of the diffusive regime (see text) from which we obtain the mean free path $l$ (the results for $l$ are listed in the figure). In the right panel one should see four slightly different full lines for different $N_c$; we show only one of them for simplicity.
 Figure (b) shows $\langle T_n \rangle$ versus $L/\xi$ for parameters indicated by bold arrows. For $n =1, 2, \dots N_c$ the resulting curves are ordered
 decreasingly.
} \label{Fig-2}
\end{figure}

 Our results are shown in figure \ref{Fig-2}. Note that the wires with grain boundaries exhibit the features typical of the white-noise-like disorder. First, $\langle \rho \rangle$ follows the usual diffusive dependence (the full line in the top left panel) in the form
\begin{equation}
 \langle \rho \rangle = 1/N_c + (2/k_F l)(L/W),
 \label{normalohm}
\end{equation}
 where $1/N_c$ is the fundamental contact resistance and the mean free path $l$ is a fitting parameter. Second, all $\langle T_n \rangle$ are equivalent in the sense that $\langle T_n \rangle \propto 1/L$ for all $n$ \cite{vanRossum}.

 The wires with rough edges exhibit a fundamentally different behavior. Specifically, the data for $\langle \rho \rangle$ follow the diffusive dependence (the full line in the top right panel) in the form
\begin{equation}
 \langle \rho \rangle = 1/N^{eff}_c + (2/k_F l)(L/W),
 \label{anomalousohm}
\end{equation}
where $1/N^{eff}_c$ is the effective contact resistance due to the $N^{eff}_c$ open channels and both $l$ and $N^{eff}_c$ are the fitting parameters. The obtained values of $N^{eff}_c$ are universal ($\simeq 6 \ll N_c$) for large $N_c$ and small $\Delta x$ (see the discussion below). The existence of the $N^{eff}_c$ open channels reflect also the transmissions in the right panel of figure \ref{Fig-2}(b). Specifically, channel $n=1$ is almost ballistic ($\langle T_1 \rangle \simeq 1$) even for $L = 0.2\xi \simeq 100 l$ and a few channels with low $n$ show $\langle T_n \rangle \sim 0.1$. Unlike the open channels, for all other channels one sees that $\langle T_n \rangle$ decays with $L$ rapidly; these channels are in the diffusive regime or even in the localization regime \cite{Feilhauer1,garcia}.

\begin{figure}[t]
\centerline{\includegraphics[clip,width=1.0\columnwidth]{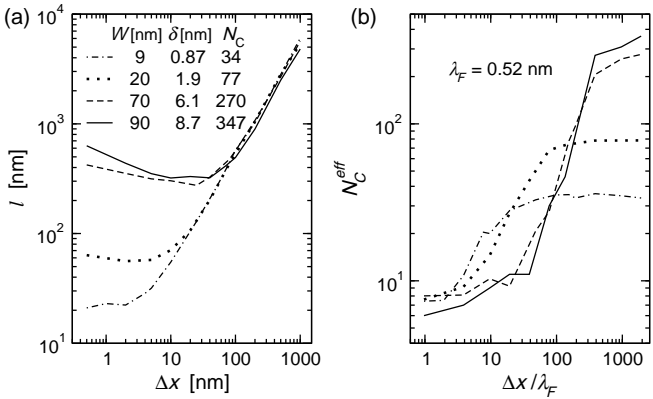}}
\vspace{-0.15cm} \caption{(a) The mean free path $l$ and (b) effective number of the open channels $N^{eff}_c$ in the wire with rough edges, both plotted in dependence on the roughness correlation length $\Delta x$ for the parameters as indicated. These data were extracted from the numerical data for
$\langle \rho \rangle$ versus $L$ by means of the fit $\langle \rho \rangle = 1/N^{eff}_c + (2/k_F l)(L/W)$, as it is explained in the text and in Fig. \ref{Fig-2}(a).
For simplicity, ratio $\delta/W$
is kept nearly the same ($\sim 1/10$) for each set of $\delta$ and $W$.} \label{Fig-3}
\vspace{-0.5cm}
\end{figure}

Figure \ref{Fig-3} shows in detail how $l$ and $N^{eff}_c$ in the wires with rough edges depend on the roughness correlation length $\Delta x$.  Indeed, the $N^{eff}_c$ versus $\Delta x$ dependence shows that $N^{eff}_c$ is a universal ($N_c$-independent) number of the order of $10$ for small enough $\Delta x$ and large enough $N_c$. The universal $N^{eff}_c$  has been discovered in Ref. \cite{Feilhauer1}, here it is demonstrated for $N_c$ as large as $347$. Further, the $l$ versus  $\Delta x$ dependence shows clearly that the minimum mean free path due to the edge roughness scattering is always a few times larger than the wire width $W$. This means that the edge roughness alone cannot explain the experimental \cite{Chand,Bluhm} observation $l \lesssim W$.
We will return to this point later on.

\subsection{C. Universality of the step-shaped-roughness model}

Before we start to discuss the rings with rough edges (next sections), we want to make an important remark.
In this paper, all our transport results for the wires/rings with rough edges are obtained for the step-shaped-roughness model in figure \ref{Fig-1}(b). We wish to point out that all these results would remain the same also for models with a smoothly varying roughness. Any smoothly varying roughness can be modeled by means of the step-shaped roughness in figure \ref{Fig-1}(b) if the latter is applied as a discretization scheme with very small and very dense steps. Using this approach, all calculations presented in this paper can be repeated in principle for any roughness model. We show below that the obtained transport results would agree with the results presented in this paper, if they are compared at the same value of $L/\xi$.

It is known for the impurity disorder \cite{Markos,Vagner} that a statistical ensemble of the macroscopically-identical mesoscopic conductors with a
microscopically-different configuration of impurities exhibits the conductance distribution which is the same (for a given value of $L/\xi$) for any choice of the impurity disorder model. The weaker the disorder the better the accord of the conductance distributions for various models.

 A similar universality (the independence on the specific model of disorder) seems to exist also when disorder is due to the rough edges. The conductance calculations in Ref. \cite{garcia}, performed for the same step-shaped-roughness model as our model in figure \ref{Fig-1}(b), give a quite similar results as the conductance calculations in paper \cite{Freilikher}, performed for the smoothly varying roughness with Gaussian-correlation function. Here we demonstrate this universality by means of the direct comparison. We calculate the conductance for the smoothly-varying roughness with Gaussian correlation (model of Ref. \cite{Freilikher}), and compare it with the conductance obtained for the step-shaped-roughness model in figure \ref{Fig-1}(b).

\begin{figure}[t!]
\centerline{\includegraphics[clip,width=0.85\columnwidth]{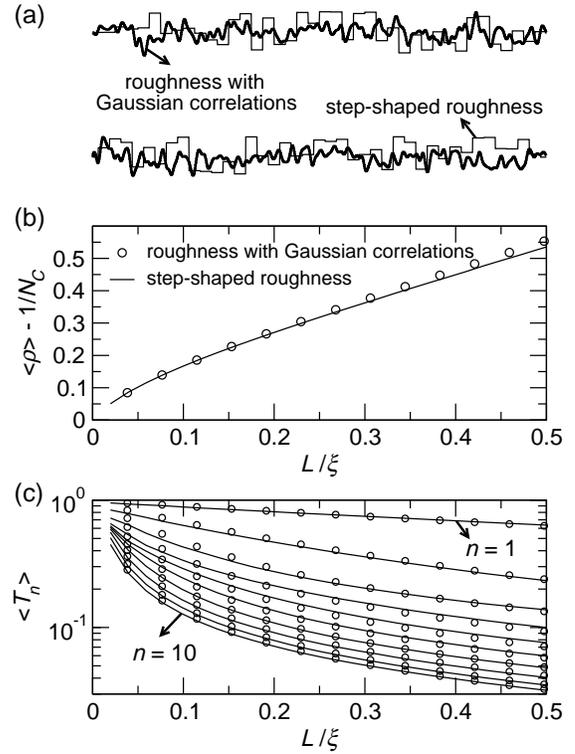}}
\vspace{-0.15cm} \caption{(a) The top view on the 2D wire with the rough edges generated numerically for two different roughness models.
In this numerical example the Au wire of width $W = 9$nm is considered, which implies that the number of the conducting channels ($N_c$) is $34$.
For the step-shaped roughness we use the RMS roughness amplitude $\delta = 0.87$nm and roughness-correlation length $\Delta x = 0.5$nm.
For the roughness with the Gaussian correlation function we choose the RMS roughness amplitude of $0.5$nm and the roughness-correlation length of $1.2$nm. In the former case we obtain the mean free path $l = 21$nm and localization length $\xi \simeq 1.4 N_c l$, and in the latter case we find $l = 20.8$nm and $\xi \simeq 1.49 N_c l$. (b) The mean resistance $\langle \rho \rangle$ versus $L/\xi$;
a comparison for the roughness models specified above. (c) The same comparative study as in figure (b), but for the channel transmissions $\langle T_n \rangle$; for clarity only the data for the first ten conducting channels are presented.
} \label{Fig-4}
\vspace{-0.5cm}
\end{figure}


In figure \ref{Fig-4} we show a typical output of our comparative study for two Au wires with the same number of the conducting channels ($N_c = 34$), so that one can compare directly the individual channel transmission. It can be seen that the individual transmissions for both roughness models are in a good agreement. This illustrates the above mentioned universality; note that the individual transmissions for both roughness models coincide albeit the values of the roughness RMS and roughness correlation length in considered roughness models are (intentionally) not the same.

 The universality exists also within the chosen roughness model. Specifically, all results of this paper and paper \cite{Feilhauer1},
 obtained for the step-shaped roughness, are the same for any choice of $\delta$ and $\Delta x$, if they are plotted in dependence on $L/\xi$.

 Finally, the main result of this paper (Fig. \ref{Fig-10}b in section IV) is that the ring with rough edges supports the ballistic persistent current $I_{typ} \simeq ev_F/L$ in spite of $L \gg l$. This results is universal simply due to its insensitivity to the edge roughness.

 \section{III. Single-electron states in clean rings: effect of Hartree-Fock interaction}

In this section we study the single-electron states in clean metal rings.
In section III.A we calculate the exact non-interacting-electron states. We point out that the ring geometry produces the centrifugal force which pushes the non-interacting states towards the outer ring edge and makes them fundamentally different from the states in the stripe geometry. In sections III.B, III.C and III.D we consider the Hartree-Fock interaction
and we find that the non-interacting-electron ring model fails. Namely, the Hartree-Fock interaction
eliminates the centrifugal force and causes that the true single-electron states in the ring are in fact similar to those ones in the stripe.

This similarity has a serious implication. We have seen in section II that the stripe with rough edges possesses a ballistic channel (channel $n=1$) even if $L \gg l$. The same has to hold for the corresponding ring. The ring with rough edges should therefore support ballistic persistent current $I_{typ} \simeq ev_F/L$ for $L \gg l$. This effect will be studied in section IV.
\begin{figure}[t]
\begin{center}
\includegraphics[clip,width=0.45\columnwidth]{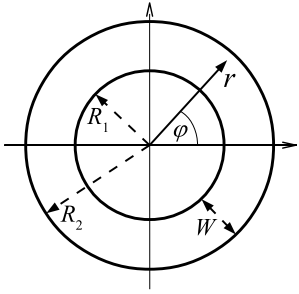}
\end{center}
\caption{The 2D ring with the inner radius $R_1$ and outer radius $R_2$. The mean radius is $R = (R_2 + R_1)/2$, the ring width $W = R_2 - R_1$. The data in the next two figures are calculated for $R_1 = 6.64$nm and $R_1 = 15.64$nm (the ring width $W = 9$nm, the ring length $L = 2 \pi R = 70$nm), and for $m^*$ equal to the free electron mass.} \label{Fig-5}
\end{figure}

\subsection{A. Clean ring with non-interacting electrons}

Consider the 2D ring (figure \ref{Fig-5}) in the form of the annulus with the inner radius $R_1$ and outer radius $R_2$.
The non-interacting electrons in the ring without magnetic flux are described by the Schrodinger equation
\begin{equation}
 H_0\psi(r,\varphi)=E\psi(r,\varphi),
\label{schrodgen 1}
\end{equation}
where $\psi$ is the wave function, $E$ is the energy, and
\begin{equation}
H_0 = - \frac{\hbar^2}{2m^*} \left( \frac{\partial^2}{\partial r^2} + \frac{1}{r}\frac{\partial}{\partial r} + \frac{1}{r^2} \frac{\partial^2}{\partial \varphi^2} \right) + V(r) .
\label{ringnoninteract}
\end{equation}
Here $r$ and $\varphi$ are the polar electron coordinates (figure \ref{Fig-5}), and $V(r)$ is the confining potential
\begin{eqnarray}
V(r) = \left\{
                          \begin{array}{ll}
                            0, & R_1<r<R_2 \\
                            \infty, & \mbox{elsewhere}
                          \end{array} \right.
,
\label{smoothpot}
\end{eqnarray}
If one sets into the equation \eqref{schrodgen 1} the wave function in the form
\begin{equation}
\label{subpasovafunkcia 1}
\psi(r,\varphi) = \frac{1}{\sqrt{L}} e^{i m \varphi} \xi(r), \quad m = 0, \pm 1, \pm 2, \dots \,
\end{equation}
where $\xi(r)$ is the radial wave function and $m$ is the angular quantum number, one obtains the radial Schrodinger equation
\begin{equation}
\left[-\frac{\hbar^2}{2m^*} \left( \frac{\partial^2}{\partial r^2} + \frac{1}{r}\frac{\partial}{\partial r} - \frac{m^2}{r^2}  \right) + V(r) \right] \xi(r)=E\xi(r)
\label{schrodgen2}
\end{equation}
This equation determines the spectrum of energies $E_{n,m}$ and wave functions $\xi_{n,m}(r)$, where $n = 1, 2, \dots$.
We obtain $E_{n,m}$ and $\xi_{n,m}(r)$ exactly by solving the equation \eqref{schrodgen2} numerically.

Magnetic flux $\Phi$  can be introduced by applying the substitution $m \rightarrow (m + \Phi/\Phi_0)$ in the Hamiltonian of equation \eqref{schrodgen2} and in the factor $e^{i m \varphi}$ of equation \eqref{subpasovafunkcia 1}. If we do so, the wave functions
$\xi_{n,m}(r)$ and $\xi_{n,-m}(r)$ are no longer degenerate. However, we find that the difference between them small and we therefore discuss only $\xi_{n,m}(r)$ calculated for $\Phi = 0$.

\begin{figure}[t]
\centerline{\includegraphics[clip,width=0.98\columnwidth]{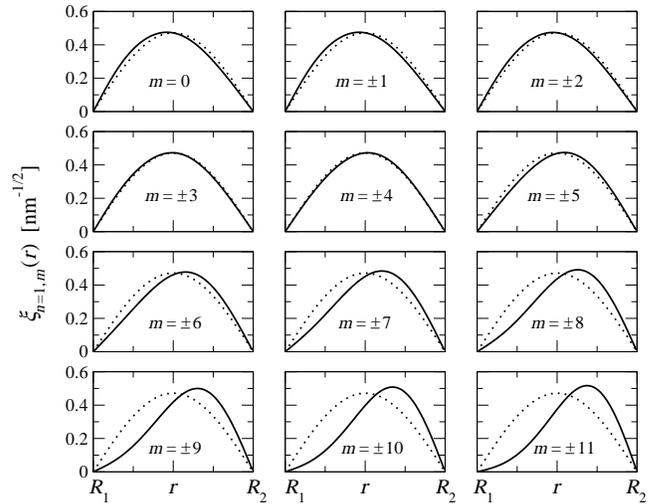}}
\vspace{-0.15cm} \caption{The full lines show the exact wave functions $\xi_{n=1,m}(r)$ of the non-interacting electrons in the ring geometry (Fig. \ref{Fig-5}). These wave functions are normalized as $(2 \pi / L) \int_{R_1}^{R_2} dr r |\xi_{n,m}(r)|^2=1$. The dotted lines show the electron wave function in the 2D stripe, $\chi_{n=1}(r) = \sqrt{2/W} \sin \left[ \frac{\pi}{W} \left( r - R_1 \right) \right]$.} \label{Fig-6}
\vspace{-0.5cm}
\end{figure}

Figure \ref{Fig-6} shows the wave functions $\xi_{n=1,m}(r)$ calculated for the ring in figure \ref{Fig-5}. They are compared with wave function $\chi_{n=1}(r) = \sqrt{2/W} \sin \left[ \frac{\pi}{W} \left( r - R_1 \right) \right]$ which holds for the stripe geometry (equation \eqref{schrodgen2}
describes the stripe geometry if the term $\frac{1}{r}\frac{\partial}{\partial r}$ is skipped and the term $- \frac{m^2}{r^2}$ is replaced by $- \frac{m^2}{R^2}$).
The difference between the electron states in the ring and electron states in the stripe is clearly visible: in the ring the function $\xi_{n=1,m}(r)$ is shifted towards the outer ring edge by the centrifugal potential $\propto \frac{m^2}{r^2}$ and towards the inner ring edge by term $\frac{1}{r}\frac{\partial}{\partial r}$. Evidently, the shift towards the outer edge dominates for large $|m|$. This shift means that the electrons in channel $n=1$ increasingly hit the outer ring edge.

A similar finding has been reported in works \cite{Jalabert,Samokhin} where
the non-interacting electron states in metallic rings were analyzed in terms of the semiclassical trajectories.
In the non-interacting model \cite{Jalabert,Samokhin} the electron wave functions are governed exclusively by the straight-line trajectories. In the annular geometry with $L \gg W$ it is clear on the first glance that any straight-line trajectory has to hit the outer ring edge many times in order to make one trip around the ring. In particular, the so-called whispering gallery modes \cite{Jalabert,Samokhin} hit solely the outer edge, in accord with our observation that $\xi_{n=1,m}(r)$ tends to be localized at $r=R_2$. As a result, one finds \cite{Samokhin} in channel $n=1$ the mean free path $l \sim W$ when the ring edges are rough. We will see that
these findings, including $l \sim W$ for $n=1$, are artefacts of the non-interacting model: they fail in the presence of the Hartree-Fock interaction.

\subsection{B. Hartree-Fock equation for clean ring}

We still consider the single-electron states in the form
\begin{equation}
\label{subpasovafunkcia 1 nm}
\psi_{n,m}(r,\varphi) = \frac{1}{\sqrt{L}} e^{i m \varphi} \xi_{n,m}(r),
\end{equation}
 however, they are now described by the Hartree-Fock equation
 \begin{eqnarray}
 [H_0 + H(r)]\psi_{nm}(r,\varphi)+F_{nm}(r,\varphi)  =
E_{nm} \psi_{nm}(r,\varphi),
\label{HFschrodgen 1}
\end{eqnarray}
where $H_0$ is the Hamiltonian of the non-interacting electrons (Eq.\ref{ringnoninteract}), $H(r)$ is the Hartree interaction
and $F_{nm}(r,\varphi)$ is the Fock interaction. The Hartree interaction reads \cite{Simanek}
 \begin{equation}
\label{hartreepot}
H(r) = - \frac{e}{4 \pi \epsilon} \int_{0}^{2 \pi} d\varphi \int_{R_1}^{R_2} dr' r' \frac{ \rho(r') }{\sqrt{r^2 + r'^2 - 2 r r' \cos{\varphi}}},
\end{equation}
where $\epsilon$ is the permittivity of the metal and \cite{Simanek}
 \begin{equation}
\label{Hartreechargedensity}
\rho(r) = -2\frac{e}{L} \sum_{n} \sum_{m} \left[ |\xi_{n,m}(r)|^2 - |\chi_n(r)|^2 \right],
\end{equation}
is the space charge density. Here we sum over all occupied states $(n,m)$, the factor of $2$ incorporates two spin orientations, and
\begin{equation}
\label{transversalstripe}
\chi_{n}(r) = \sqrt{2/W} \sin \left[n \pi \left( r - R_1 \right)/W \right]
\end{equation}
is the wave function in the stripe. The charge density \eqref{Hartreechargedensity} is due to the ring geometry: if we skip in equation \eqref{schrodgen2} the term $\frac{1}{r}\frac{\partial}{\partial r}$ and replace the term $- \frac{m^2}{r^2}$ by $- \frac{m^2}{R^2}$, we obtain the stripe geometry with solution $\xi_{n,m}(r) \equiv \chi_n(r)$ and $\rho(r)=0$. In golden rings there are many occupied channels and the term
$2\frac{e}{L} \sum_{n} \sum_{m} |\chi_n(r)|^2$ in \eqref{Hartreechargedensity} is equal to the charge density of the positive ion background.
Finally, the Fock interaction is operative between the electrons of like spin. It reads
\begin{eqnarray}
\nonumber
&& F_{nm}(\varphi, r)=
\\
 &-& \nonumber
 \frac{e^2}{4 \pi \epsilon}  \int^{R_2}_{R_1} d r' r' \int^{2 \pi}_{0} d\varphi '
\frac{ (1/\sqrt{L}) e^{i m \varphi'} \xi_{nm} (r')}{\sqrt{ r^2 + r'^2 - 2rr'\cos(\varphi - \varphi') }}
 \\
 &\times&
 \frac{1}{L} \sum_{n'} \sum_{m'} e^{i m' (\varphi - \varphi')} \xi_{n' m'}(r) \xi_{n' m'}(r'),
\label{potk}
\end{eqnarray}
where we sum over all occupied states $(n', m')$. Evidently,
\begin{equation}
F_{nm}(\varphi, r) = \frac{1}{\sqrt{L}} e^{i m \varphi} F_{nm}(r),
\label{Fockangleradial}
\end{equation}
where $F_{nm}(r)$ is the radial part of $F_{nm}(\varphi, r)$:
\begin{eqnarray}
\nonumber
&& F_{nm}(r) =
\\
&-&  \nonumber
\frac{e^2}{4 \pi \epsilon} \frac{1}{L}  \int^{R_2}_{R_1}  d r' r' \int^{2 \pi}_{0} d\theta
\frac{e^{-i m \theta} \xi_{nm} (r')}{\sqrt{ r^2 + r'^2 - 2rr'\cos(\theta) }}
\\
 &\times&
\sum_{n'} \sum_{m'} e^{i m' \theta} \xi_{n' m'}(r) \xi_{n' m'}(r').
\label{Fockradial}
\end{eqnarray}
If we set into Eq. \eqref{HFschrodgen 1}
equations  \eqref{subpasovafunkcia 1 nm} and \eqref{Fockangleradial}, we obtain
\begin{eqnarray}
\nonumber
&& \left[-\frac{\hbar^2}{2m^*} \left( \frac{\partial^2}{\partial r^2} + \frac{1}{r}\frac{\partial}{\partial r} - \frac{m^2}{r^2}  \right) + V(r)
+ H(r)  \right] \xi_{n,m}(r)
\\
&&
 +  F_{nm}(r) = E_{n,m}\xi_{n,m}(r) .
\label{schrodgen2Hartree}
\end{eqnarray}
The last equation is the radial Hartree-Fock equation.

Equation \eqref{schrodgen2Hartree} can be solved numerically by means of the Hartree-Fock iterations.
In the first iteration step, the Hartree term (eqs. \ref{hartreepot} and \ref{Hartreechargedensity}) and Fock term \eqref{Fockradial} are calculated by setting for $\xi_{nm} (r)$  the exact non-interacting ring states and equation \eqref{schrodgen2Hartree} is solved numerically. This gives a new set of states $\xi_{nm} (r)$. In the second iteration step, the Hartree term and Fock term are calculated for $\xi_{nm} (r)$ obtained in the first iteration step and equation \eqref{schrodgen2Hartree} is solved again. After many iterations a self-consistent solution is achieved;
the wave functions $\xi_{nm} (r)$ obtained in two successive steps show a negligible difference. Since the self-consistent calculation is computationally cost, in this paper we obtain the self-consistent Hartree-Fock results only for rings with a single occupied channel. For the
multi-channel rings we perform either only the first Hartree-Fock iteration step or a so-called restricted self-consistent Hartree-Fock calculation.
In spite of these restrictions, we are able to draw a few key conclusions.

\begin{figure}[t]
\begin{center}
\includegraphics[clip,width=0.98\columnwidth]{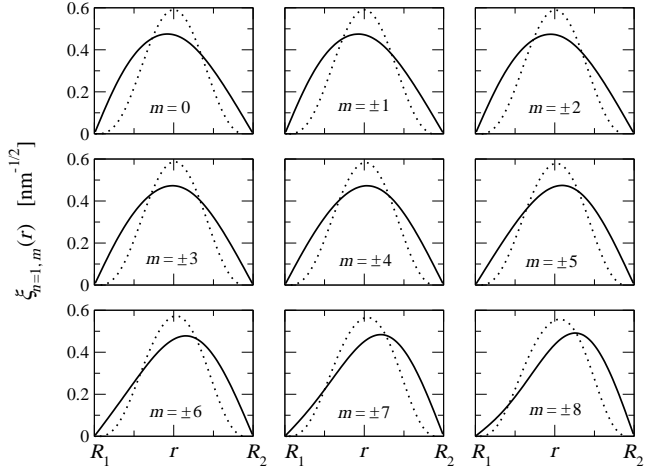}
\end{center}
\caption{Electron wave functions $\xi_{n=1,m}(r)$ for the ring in Fig. \ref{Fig-5}. The full lines show the results for the non-interacting electrons (taken from the preceding figure) and the dotted lines show the self-consistent results for the electrons that interact via the Hartree-Fock interaction. These Hartree-Fock calculations were performed for $34$ electrons in the ground-state $n=1, m=0, \pm 1, \pm 2, \dots, \pm 8$; each ($n,m$) is occupied by two electrons with opposite spins. The permittivity $\epsilon$ is assumed to be equal to the permittivity of vacuum. } \label{Fig-7}
\end{figure}

\subsection{C. Hartree-Fock results: failure of the non-interacting model}

Figure \ref{Fig-7} shows again the wave functions $\xi_{n=1,m}(r)$ for the ring in figure \ref{Fig-5}. The full lines show the exact non-interacting ring states (taken from the preceding figure), the dotted lines show the self-consistent Hartree-Fock results. For simplicity, in this Hartree-Fock calculation the electron number in the ring is restricted to $34$ in order to occupy only channel $n=1$.
The results clearly illustrate why the non-interacting model fails. As $|m|$ increases, the non-interacting electron states (full lines) are pushed by the centrifugal force towards the outer ring edge. However, the Hartree-Fock interaction repels the electrons back.
The Hartree-Fock wave functions are almost symmetric around the center of the ring cross section even for large $|m|$. Furthermore, this symmetric shape is so narrow that the wave-function tails do not reach the ring edges. This implies that the electrons in channel $n =1$ move
 around the ring ballistically - without collisions with the ring edges. This however also means that channel $n=1$ will be ballistic even if the ring edges are rough, similarly as we have seen for the stripe geometry (the bottom right panel of figure \ref{Fig-2}).

     Figure \ref{Fig-7} also suggests that the true single-electron states of the clean ring, the Hartree-Fock states $\xi_{n=1,m}(r)$, can be well approximated by the non-interacting-electron wave-function of the clean stripe,
$\chi_{n=1}(r) = \sqrt{2/W} \sin \left[\frac{\pi}{W} \left( r - R_1 \right) \right]$. Clearly, the function $\chi_{n=1}(r)$
 captures the fact that the effect of the ring curvature is compensated by the Hartree-Fock field. Additionally, it is not as narrow as the Hartree-Fock states $\xi_{n=1,m}(r)$ and thus suppresses the collisions with the ring edges less effectively (one does not need to worry that the suppression is overestimated). Unlike $\chi_{n=1}(r)$, the exact non-interacting ring states $\xi_{n=1,m}(r)$ evidently fail to mimic the true single-electron ring states.  Now we show that these findings hold also for the multi-channel rings.

\subsection{D. Hartree-Fock results continued: multi-channel rings}

\begin{figure}[t]
\begin{center}
\includegraphics[clip,width=0.94\columnwidth]{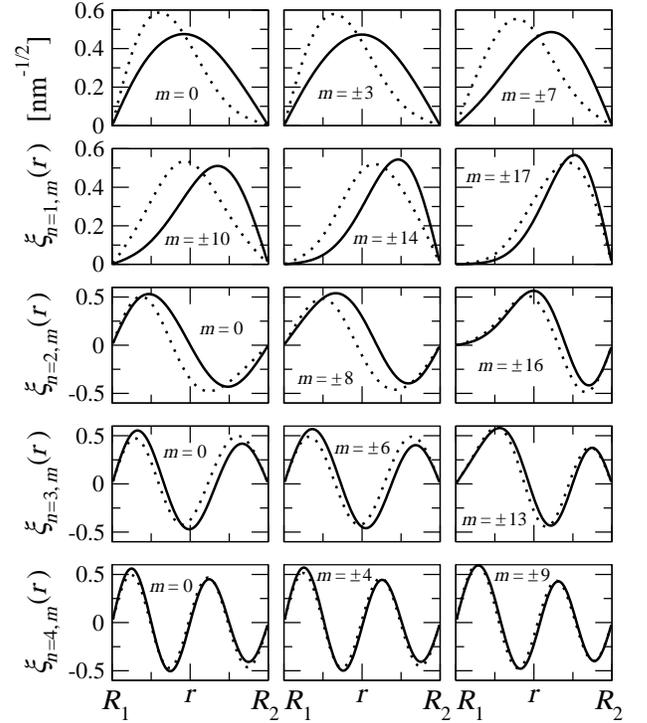}
\end{center}
\caption{Electron wave functions $\xi_{n=1,m}(r)$, $\xi_{n=2,m}(r)$, $\xi_{n=3,m}(r)$, and $\xi_{n=4,m}(r)$ for the ring in figure \ref{Fig-5}. The ring is filled by $228$ electrons, these electrons occupy four channels, each occupied state $(n,m)$ contains two electrons with opposite spins. Specifically, in channel $n=1$ there are $70$ electrons in states $m=0, \pm 1, \dots, \pm 17$, in
channel $n=2$ there are $66$ electrons in states $m=0, \pm 1, \dots, \pm 16$, channel $n=3$ contains $54$ electrons in states $m=0, \pm 1, \dots, \pm 13$, and channel $n=4$ contains $38$ electrons in states $m=0, \pm 1, \dots, \pm 9$. The figure shows the results for selected values of $m$. The full lines are the results for the non-interacting electrons (obtained by solving equation \ref{schrodgen2}). The dotted lines are the Hartree-Fock results due the first Hartree-Fock iteration step.} \label{Fig-8}
\end{figure}

 A golden 2D ring of size considered in figure \ref{Fig-5} contains about thousand electrons which occupy about thirty channels $n$. A self-consistent Hartree-Fock analysis of such many-electron ring is beyond our computational possibilities.
However,
useful information can be obtained already when only the first Hartree-Fock iteration is performed for the ring with a few occupied channels.
Results of such calculation are shown in figure \ref{Fig-8} for the ring with four occupied channels. The full lines show the exact non-interacting ring states and the dotted lines show the Hartree-Fock states due to the first iteration. The following features are worth noticing.

  As before, the exact non-interacting states are pushed towards the outer ring edge by centrifugal force, while the Hartree-Fock interaction repels the electrons in the opposite direction.
  In particular, most of the Hartree-Fock wave functions in channel $n=1$ is now shifted towards the inner edge rather than towards the outer edge.
  Thus, the key feature of the exact non-interacting states (the strong shift towards the outer edge by the centrifugal force) tends to diminish
  when the states are subjected to their own Hartree-Fock field.
  This is a clear sign that the exact non-interacting states fail to describe the true single-electron states in metallic rings. This also means that the modeling of the single-electron states in metallic rings by means of the straight-line paths \cite{Jalabert} fails for real metal rings: the electron paths in presence of the Hartree-Fock field cannot be the straight lines.

  Figure \ref{Fig-8} also shows that the Hartree-Fock states approach the non-interacting states as the channel number $n$ increases. Indeed, with the increase of $n$ the centrifugal term $\propto m^2$ becomes less important because the larger the number $n$ the smaller the occupied angular numbers $m$ in channel $n$. As a result, the exact non-interacting states approach the stripe-geometry limit $\chi_{n}(r) = \sqrt{2/W} \sin \left[n \frac{\pi}{W} \left( r - R_1 \right) \right]$, and become robust against the Hartree-Fock field.

In summary, the first Hartree-Fock iteration step in figure \ref{Fig-8} shows that the exact non-interacting electron model fails to describe the true single-electron states in a clean multi-channel metal ring. We have performed a similar first-iteration-step calculation also for three other rings with the same size but with a larger electron number: $6$, $9$, and $17$ occupied channels. We have seen a clear trend: the larger the electron number, the stronger the shift of the non-interacting states towards the outer ring edge and the larger the opposite-oriented shift of the Hartree-Fock states. In other words, with increasing Fermi energy the difference between the non-interacting states and Hartree-Fock states increases and the failure of the non-interacting-ring model is more pronounced.

What are the true self-consistent Hartree-Fock states in multi-channel rings? The multi-channel self-consistent calculation is for us too cost; a feasible task is the \emph{restricted self-consistent} Hartree-Fock calculation. This means that we calculate the wave functions $\xi_{n,m}(r)$ self-consistently for one selected channel (say channel $n=1$) by assuming that the electrons in channel $n=1$ interact with the self-consistent Hartree-Fock potential due to the electrons in channel $n=1$ and with the non-self-consistent Hartree-Fock potential due to the electrons in channels $n=2,3 \dots$. The \emph{non-self-consistent} means that the Hartree-Fock potential due to channels $n=2,3, \dots$ is calculated by setting for $\xi_{n=2,m}(r)$, $\xi_{n=3,m}(r)$, $\dots$ the exact non-interacting states rather than the self-consistent Hartree-Fock states.

\begin{figure}[t]
\begin{center}
\includegraphics[clip,width=0.74\columnwidth]{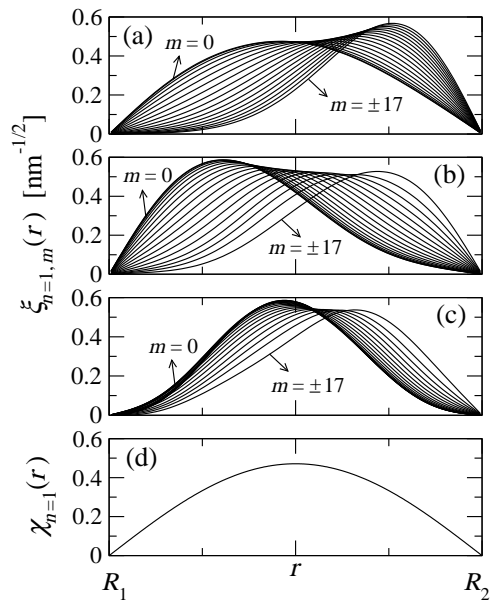}
\end{center}
\caption{Electron wave functions $\xi_{n=1,m}(r)$ with $m = 0, \pm 1, \dots \pm 17$ in the ring with four occupied channels, considered in Fig. \ref{Fig-8}. Results from figure \ref{Fig-8} are shown again in panels (a) and (b), where panel (a) shows the exact non-interacting wave functions and panel (b) shows the Hartree-Fock wave functions due to the first iteration step. Panel (c) shows the results of the \emph{restricted self-consistent} Hartree-Fock calculation (see the main text).
Panel (d) shows the stripe-geometry solution $\chi_{n=1}(r) = \sqrt{2/W} \sin \left[\frac{\pi}{W} \left( r - R_1 \right) \right]$.} \label{Fig-9}
\end{figure}

In figure \ref{Fig-9} we show again the non-interacting electron states (panel a) and Hartree-Fock states from the first iteration step (panel b), and we compare them with results of the restricted self-consistent Hartree-Fock calculation (panel c). Unlike the non-interacting states in panel a, the Hartree-Fock wave functions in panel c are repelled back to the center. Moreover, when compared with the wave functions in panels a and b, the wave functions in panel c show a tendency to be compressed to the same symmetric form. This tendency suggests that
the fully self-consistent Hartree-Fock procedure would make the wave functions in panel c even more symmetric and even closer to each other.
Another support for this suggestion is provided by the single-channel-ring study in figure \ref{Fig-7}, where
the (fully self-consistent) Hartree-Fock states $\xi_{n=1,m}(r)$ are indeed almost the same and almost perfectly symmetric.
So we believe that the stripe-geometry limit $\chi_{n=1}(r) = \sqrt{2/W} \sin \left[\frac{\pi}{W} \left( r - R_1 \right) \right]$, proposed above for the single-channel rings, approximates well also the true single-electron states $\xi_{n=1,m}(r)$ in multi-channel rings.

 Note that $\chi_{n=1}(r)$ approximates quite well already the (not fully self-consistent) results in panel c. First, it captures the tendency of the Hartree-Fock interaction to compensate the effect of the ring curvature. Second, one sees in panel c that the wave function tails near the edge points $R_1$ and $R_2$ are mostly suppressed much more than the tails of $\chi_{n=1}(r)$. Therefore, the Hartree-Fock states in panel c have to feel the edge roughness (if any) less efficiently than it is felt by state $\chi_{n=1}(r)$. Thus, approximation $\xi_{n=1,m}(r) \simeq \chi_{n=1}(r)$ certainly does not underestimate the edge roughness scattering in the ring.

 Finally, approximation $\xi_{n=1,m}(r) \simeq \chi_{n=1}(r)$ can be extended to all $n$ as $\xi_{n,m}(r) \simeq \chi_{n}(r)$, because
  the effect of the centrifugal force diminishes  with increasing $n$ (figure \ref{Fig-8}). In conclusion, the true single-electron states of the clean metallic ring, the self-consistent Hartree-Fock states, can be approximated by the non-interacting states of the clean metallic stripe,
\begin{equation}
\label{subpasovafunkcia 1 nm approx}
\psi_{n,m}(r,\varphi) \simeq \frac{1}{\sqrt{L}} e^{i m \varphi} \sqrt{\frac{2}{W}} \sin \left[n\frac{\pi}{W} \left( r - R_1 \right) \right],
\end{equation}
of course, with eigen-energies
\begin{equation}
\label{subpasovaenergia 1 nm approx}
E_{n,m} \simeq \frac{\hbar^2 \pi^2}{2m^*W^2} n^2 + \frac{\hbar^2}{2m^*R^2} m^2.
\end{equation}
To add magnetic flux $\Phi$, substitution $m \rightarrow (m + \Phi/\Phi_0)$ has to be used on the right-hand side of equations \eqref{subpasovafunkcia 1 nm approx} and \eqref{subpasovaenergia 1 nm approx}. Approximation $\xi_{n,m}(r) \simeq \chi_{n}(r)$
captures the fact that the effect of $\Phi$ on the numerically obtained $\xi_{n,m}(r)$ is hardly visible.

We have seen in section II that in the stripe with rough edges the channel $n=1$ is ballistic for $L \gg l$.
Since $\xi_{n=1,m}(r) \simeq \chi_{n=1}(r)$, channel $n=1$ has to be ballistic also in the ring with rough edges, and such ring
  should therefore support ballistic persistent current $I_{typ} \simeq ev_F/L$ even if $L \gg l$. This ballistic current is studied in the next section. In contrast to our result, the non-interacting model predicts \cite{Jalabert,Samokhin} for channel $n=1$ the diffusive mean free path $l \sim W$, whenever the ring with rough edges is of size $L \gg W$. This prediction is an artefact of the non-interacting model.

 \section{IV. Persistent currents in rings with grain boundaries and rough edges}

Assume that the wires in figures \ref{Fig-1}a and \ref{Fig-1}b are circularly shaped in the plane of the 2D gas and the wire ends are connected. So we have a 2D ring with grain boundaries and a 2D ring with rough edges. \emph{What are the persistent currents in such rings?} In this section we answer the question by means of simple intuitive arguments (section IV.A) and by means of the first-principle simulation (section IV.B). Simulation results for typical persistent currents are presented in section IV.C, section IV.D presents the sample-specific currents. In section IV.E we simulate typical persistent currents in rings with combined disorder due to the rough edges and grain boundaries, compare them with experiment \cite{Chand}, and explain the
anomalous experimental data.

\subsection{A. Intuitive arguments}

For rings with random grain boundaries one can safely expect the standard diffusive result $I_{typ}\simeq (e v_F/L)(l/L)$, because the corresponding metallic stripe shows the standard diffusive resistance (left panels of figure \ref{Fig-2}). This expectation agrees with our microscopic results shown later. We note that \emph{our grain-boundary model (figure \ref{Fig-1}a) is universal in the sense that any other grain-boundary model with random boundaries would give again the diffusive conductance and diffusive persistent current.} Indeed, diffusive transport is caused by the random orientation and random positions of grain boundaries, not by microscopic details of the individual boundary.

For the rings with rough edges the situation is different. We have seen in section III that the electron states in clean rings and clean stripes are similar, in particular $\xi_{n=1,m}(r) \simeq \chi_{n=1}(r)$. In addition, in section II we have seen that channel $n=1$ in the stripe with rough edges possesses at $L \gg l$ the transmission $\langle T_1 \rangle \simeq 1$ (the right panel of figure \ref{Fig-2}b). Since $\xi_{n=1,m}(r) \simeq \chi_{n=1}(r)$, channel $n=1$ has to be ballistic also in the ring with rough edges and the persistent current in such ring can be estimated as follows. Assume roughly that $\langle T_n \rangle = 1$ for $n=1$
and $\langle T_n \rangle \sim l/L$ for all other $n$. In this model, channel $n=1$ contributes by ballistic
current $I_{typ} = e v_F/L$ while the total contribution from other channels is diffusive, $I_{typ} \simeq (e v_F/L)(l/L)$, and negligible for $L \gg l$. Thus, multichannel rings with rough edges should support at $L \gg l$ the typical
currents $I_{typ} \simeq e v_F/L$, expected to exist only in ballistic single-channel rings.

In terms of classical paths, the rough edges scatter all electrons except for a small part of those that move (almost) in parallel with the edges. This small part, mainly the electrons that occupy
channel $n=1$, hits the edges rarely and thus moves almost ballistically. We recall (see section III) that the motion parallel with the edges exists in the ring geometry owing to the Hartree-Fock interaction. It eliminates the effect of the ring geometry and establishes relation $\xi_{n=1,m}(r) \simeq \chi_{n=1}(r)$.

 \subsection{B. Microscopic model}

  We start with the clean ring. According to section III, the true single-electron states of the clean ring (the Hartree-Fock states) can be approximated by the non-interacting electron states of the clean stripe as show equations \eqref{subpasovafunkcia 1 nm approx} and \eqref{subpasovaenergia 1 nm approx}. One can define variables $x$ and $y$ by transformation
 $R \varphi \rightarrow x$ and $(r - R_1) \rightarrow y$, and rewrite equations \eqref{subpasovafunkcia 1 nm approx} and \eqref{subpasovaenergia 1 nm approx} as
\begin{equation}
\label{subpasovafunkcia 1 nm ringstripe}
\psi_{n,m}(x,y) = \frac{1}{\sqrt{L}} e^{i k_m x} \sqrt{\frac{2}{W}} \sin \left[n\frac{\pi}{W} y \right]
\end{equation}
and
\begin{equation}
\label{subpasovaenergia 1 nm ringstripe}
E_{n,m} = \frac{\hbar^2 \pi^2}{2m^*W^2} n^2 + \frac{\hbar^2}{2m^*L^2} k^2_m,
\end{equation}
where
$k_m = \frac{2 \pi}{L}(m + \Phi/\Phi_0)$. If we take the Hamiltonian of the clean stripe (Hamiltonian \eqref{hamiltdisord} without disorder) and write Schrodinger equation $H \psi(x,y)= E \psi(x,y)$, the wave functions \eqref{subpasovafunkcia 1 nm ringstripe} and eigen-energies \eqref{subpasovaenergia 1 nm ringstripe} are evidently its solutions. It is customary to view this approach as a quasi-1D approximation
in which the \emph{non-interacting-electron} states of the 2D ring are naively mapped on the non-interacting electron states of the straight stripe
via transformation $R \varphi \rightarrow x$, $(r - R_1) \rightarrow y$. In fact, \emph{this mapping is not a quasi-1D approximation for the non-interacting 2D states. The states mapped on the non-interacting states of the stripe are the Hartre-Fock states of the ring, and this mapping is due to the fact that the Hartree-Fock interaction acts against the centrifugal force and eliminates the effect of the ring geometry.} If one uses this mapping, one in fact captures the key effect of the Hartree-Fock interaction without any Hartree-Fock calculation.

In case of disordered rings, the Hartree-Fock interaction is expected to play a key role in the rings with rough edges (see the discussion in section III). In this case the Hartree-Fock analysis would be even more tedious than for the clean rings. Fortunately, the mapping approach is a reasonable and viable alternative which can easy be extended to disordered rings.

 We bend the disordered 2D stripe in figure \ref{Fig-1} to form a 2D ring similar to that one in figure \ref{Fig-5}, but disordered. We describe the electron states in the ring by Hamiltonian of the constituting stripe, by Hamiltonian \eqref{hamiltdisord}. The ring is mapped on the stripe by assuming that the $x$ coordinate in Hamiltonian \eqref{hamiltdisord} is the electron position along the ring circumference
and $y$ is the position along the ring radius. We can thus apply directly the scattering matrix calculation for the disordered stripe (section II). Of course, now this calculation has to be supplemented by cyclic boundary conditions \cite{Cheung}
\vspace{-0.1cm}
\begin{equation}
\vspace{-0.1cm}
\begin{array}{c}
\psi(0,y) = \exp(- i 2 \pi \Phi/\Phi_0) \psi(L,y), \\
 \frac{\partial \psi}{\partial x}(0,y) = \exp(- i 2 \pi \Phi/\Phi_0)  \frac{\partial \psi}{\partial x}(L,y),
\end{array}
\label{hrpo}
\end{equation}
where the exponential factor is the Peierls phase. We set into equations \eqref{hrpo} the expansion \eqref{rozvoje}  and  rewrite them as
\begin{equation}
\left(
\begin{array}{c}
\vc{A}^-(0) \\
\vc{B}^+(L) \\
\end{array}
\right)
=
\left[
\begin{array}{cc}
0 & Q^{-1}(\phi) \\
Q(\phi) & 0 \\
\end{array}
\right]
\left(
\begin{array}{c}
\vc{A}^+(0) \\
\vc{B}^-(L) \\
\end{array}
\right),
\label{okrajka}
\end{equation}
where $Q$ is the $N \times N$ matrix with terms $Q_{\alpha \beta} = e^{i 2\pi \Phi/\Phi_{0}} \delta_{\alpha \beta}$. The scattering matrix equation \eqref{Smatrixrovnica} has to be fulfilled together with cyclic conditions \eqref{okrajka}. This happens for discrete energies $E =E_j(\Phi) $ which we find for a given ring numerically \cite{Feilhauer2}.

Again, it is tempting to consider the above mapping approach as a quasi-1D approximation \cite{Cheung} and to think about a \emph{truly-2D} calculation for \emph{non-interacting-electrons} [with disorder introduced in the 2D-ring Hamiltonian \eqref{ringnoninteract}]. We recall that the truly-2D calculation without the Hartree-Fock interaction fails to describe the true single-electron states in clean rings and rings with rough edges. The mapping approach captures the key effect of the Hartree-Fock interaction.

  Once we know the ring spectrum $E_j(\Phi)$, we calculate the sample-specific current $I = - {\sum}_{ \forall E_j \leq E_F} dE_j/d\Phi$ and eventually the typical current $I_{typ} \equiv \langle I^2\rangle ^{1/2}$, where $\langle I^2\rangle$ is averaged over a small energy window at $E_F$. Technical details of averaging are explained in \cite{Feilhauer2} and also in section IV.D.

\begin{figure}[t!]
\centerline{\includegraphics[clip,width=0.97\columnwidth]{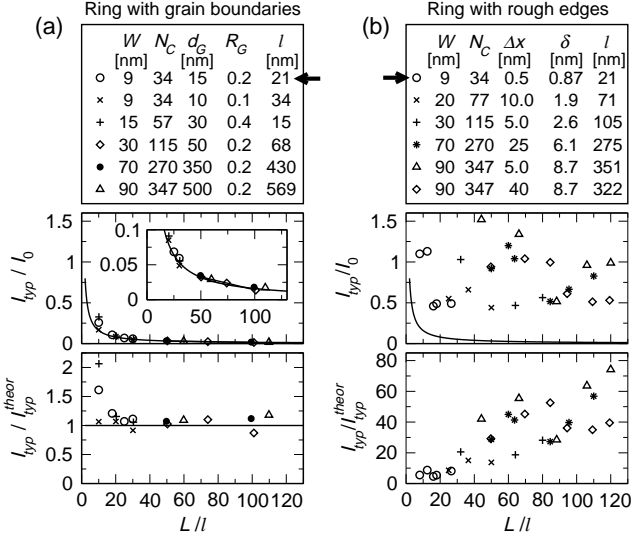}}
\caption{Typical persistent current $I_{typ}$ versus $L/l$ in disordered Au ring. The ring parameters are shown, $\Phi=-0.25 h/e$, $l$ has been obtained from the wire resistivity (figure \ref{Fig-2}). The arrows
point the parameters studied further in figure \ref{Fig-12}. Symbols are our data, full lines show formula $I^{theor}_{typ} = 1.6 (e v_F/L) (l/L)$.} \label{Fig-10}
\end{figure}

 \subsection{C. Results for typical currents: a comparison for random grain boundaries and rough edges}

Figure \ref{Fig-10} shows our main results.
For the rings with grain boundaries one can see that our data for $I_{typ}$ agree (at large $L$) with the diffusive result $I^{theor}_{typ} = 1.6 (e v_F/L) (l/L)$. This agrees with experiments \cite{Bluhm,Bles}, illustrates the universality (the white-noise-like properties) of our random-grain-boundary model, and confirms the intuitive expectations of section IV.A.

For
the rings with rough edges, however, our data for $I_{typ}$ are systematically (not regarding the data fluctuations) close to the ballistic one-channel value $I_0 = e v_F/L$, albeit $L \gg l$, $N_c \gg 1$, and
$\langle \rho \rangle \propto L$. All this agrees with experiment \cite{Chand,Riedel} and this agreement is discussed in detail in section IV.E. In the preceding text we have arrived at result $I_{typ} \sim ev_F/L$ intuitively by assuming, that the electrons in channel $n=1$ almost entirely avoid the scattering with rough edges and thus carry the ballistic current $\sim e v_F/L$. Now we make this intuitive argument more precise.

\begin{figure}[t!]
\centerline{\includegraphics[clip,width=1.0\columnwidth]{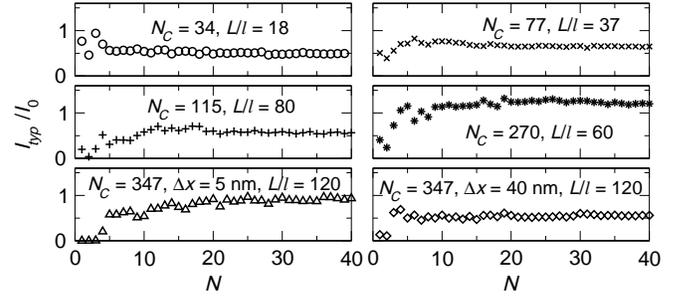}}
\vspace{-0.15cm} \caption{Typical persistent current $I_{typ}$ in the ring with rough edges as a function of the total number of channels ($N$) considered in the simulation. The same parameters and symbols are used as
in figure \ref{Fig-10}(b), the considered ring lengths are shown as $L/l$.} \label{Fig-11}
\vspace{-0.5cm}
\end{figure}

In figure \ref{Fig-11} we show how the typical current in the ring with rough edges depends on the number of
channels ($N$) considered in the simulation. It is (roughly) $N$-independent for $N \gtrsim 10$, no matter how large $N_c$ is. In other words, the currents $\sim I_0$ in rings with rough edges exist owing to the open channels $n = 1, 2, \dots, N^{eff}_c$, where $N^{eff}_c \sim 10$ for any value of $N_c$ (see also figure \ref{Fig-3}).  Since $\langle T_1 \rangle \sim 1$, our intuitive argument invokes that the value $I_{typ} \sim I_0$ will survive also if one chooses $N$ as small as $N=1$. Figure \ref{Fig-11} shows that this is not the case. For instance, in the ring with $N_c = 347$ and $L/l = 120$ the current for $N \rightarrow 1$ is quite close to zero. This is easy to understand: Once the channel $n = 1$ cannot communicate with other channels,
 the transmission $\langle T_1 \rangle \sim 1$ tends to be suppressed to zero by Anderson localization, present in any sufficiently long 1D disordered system. \emph{Communication with a few other channels is needed to restore $\langle T_1 \rangle \sim 1$ and to obtain $I_{typ} \sim I_0$.}

\begin{figure}[t!]
\centerline{\includegraphics[clip,width=1.0\columnwidth]{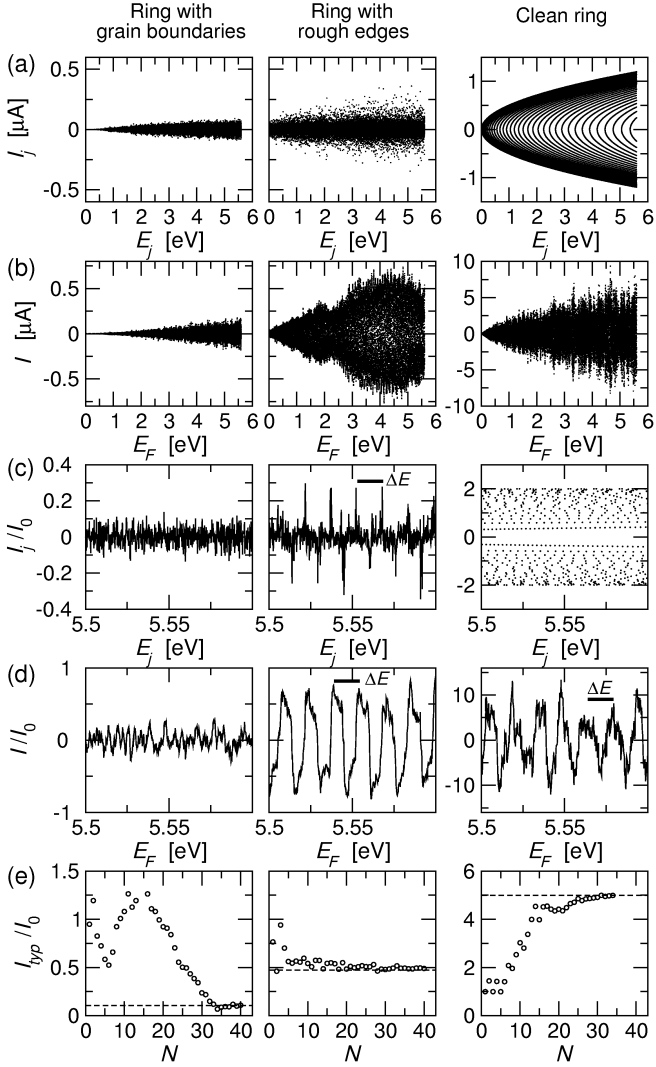}}
\vspace{-0.15cm} \caption{Persistent currents in a ring with grain boundaries, a ring with rough edges, and a clean ring for the parameters marked by the arrows in Fig.~\ref{Fig-10}, for $L=375$ nm, and for $\Phi=-0.25 h/e$. For both disordered rings, the considered parameters ensure
$l(E_F)=21$ nm at the Au Fermi level ($E_F = 5.6$ eV). Figure (a) shows the single-electron current $I_j$ versus the eigen-energy $E_j$. Figure (b) shows the total current $I = \sum_{ \forall E_j \leq E_F} I_j$ obtained by summing the currents in the figure (a) for
$E_F$ varied from $0$ to $5.6$ eV. Figures (c) and (d) show the same data as the figures (a) and (b), but for a small energy window below the Au Fermi level. The data are scaled by $I_0= e v_F/L$,
the data points are connected by full lines which serve as a guide for the eye, the bars depict the energy increment $\Delta E = 2 \pi \hbar v_F /L$.
Figure (e) shows the typical current $I_{typ} \equiv \langle I^2\rangle ^{1/2}$.
Averaging over the energy window in figure (d) gives the values shown by dashed lines:
$I_{typ}/I_0 \simeq 1.6 (l/L)$ for the ring with grain boundaries, $I_{typ}/I_0 \simeq 0.5$ for
the ring with rough edges, and $I_{typ}/I_0 \simeq \sqrt{N_c}$
for the clean ring \cite{IBM}. The circles show the data obtained by varying
the number of channels, $N$, from $N=1$ to $N > N_c$ (here $N_c = 34$). } \label{Fig-12}
\vspace{-0.5cm}
\end{figure}

\subsection{D. Sample-specific currents}

To provide further insight, figure \ref{Fig-12} shows the sample-specific currents in two selected rings from figure \ref{Fig-10} (bold arrows) and in a clean ring. Figure \ref{Fig-12}(a) shows the dependence $I_j$ versus $E_j$, figure \ref{Fig-12}(b) shows the total current $I = \sum_{ \forall E_j \leq E_F} I_j$ versus $E_F$.
 Evidently, the ring with rough edges exhibits remarkably larger currents than the ring with grain boundaries, albeit both rings are of the same size and posses the same value of $l$.

 Figures \ref{Fig-12}(c) and \ref{Fig-12}(d) focus on a small energy window below the Au Fermi level. One can see that $I_j$ in the ring with rough edges exhibits sharp peaks with the sign alternating and oscillating with period
 $\Delta E = 2 \pi \hbar v_F /L$. This period is twice the inter-level distance in the ballistic single-channel ring,
 which suggests that the peaks are due to the quasi-ballistic channel $n=1$. [We recall that $\langle T_1 \rangle \sim 1$ also for $L/l \gg 1$, as is shown in the right panel of figure \ref{Fig-2}(b).] However, the height of the peaks is affected also by other channels, because, as discussed above, channel $1$ cannot keep $\langle T_1 \rangle \sim 1$ without communicating with a few other channels.

 In figure \ref{Fig-12}(d) one can see that in the ring with rough edges
 also the total current $I(E_F)$
 oscillates with period $\Delta E$. The amplitudes of the total current are close to $I_0$, and therefore the typical currents
 of size $\sim I_0$ appear in figure \ref{Fig-10}(b).

In fact, already the data for the clean ring
show $I(E_F)$ oscillating with period
$\Delta E$. However, the amplitude of $I$ is $\sim \sqrt{N_c}2I_0$ \cite{IBM} and the amplitude of $I_n$ is $2I_0$, where
the factor of $2$ is due to the spin.
Evidently, the rough edges reduce $I$ from $\sim \sqrt{N_c}2I_0$ to $\sim I_0$, but they do not change the oscillation period set by the clean ring. Note that also the ring with grain boundaries exhibits the oscillating persistent current. These oscillations are chaotic and correlated with correlation length $\sim (l/L) \Delta E$, predicted \cite{Cheung,Riedel} for the white-noise-like disorder.

Figure \ref{Fig-12}(e) shows the typical current.
The dashed lines show the values of $I_{typ}$ obtained from the data in figure \ref{Fig-12}(d), the circles
show $I_{typ}$ in dependence on $N$.  For all three rings one sees, that the circles approach with raising $N$ the $N$-independent value (the large $N$ limit)
represented by the dashed line. It can be seen that a reliable estimate of $I_{typ}$ in the ring with grain boundaries requires $N \gtrsim N_c$, while for the ring with rough edges one only needs $N \sim 10$ no matter how large $N_c$ is. This is due to the effective number $N^{eff}_c \sim 10$, as
has already been explained in the beginning of this section.

\subsection{E. Combined effect of rough edges and random grains: Comparison with experiment}

In experiment \cite{Chand}
the persistent current $\sim I_0$ was observed
in the Au ring with $L \simeq 100 l$ and $W = 90$nm.
Indeed, figure \ref{Fig-10}(b) demonstrates $I_{typ} \sim I_0$ also for $L/l \simeq 100$ and $W = 90$nm. The difference is that the work \cite{Chand} has reported $l \simeq W$ ($l = 70$nm for $W = 90$nm)
while our values of $l$ in Fig. \ref{Fig-10}(b) [see also figure \ref{Fig-3}(a)] are
at least two times larger than $W$; the edge roughness alone cannot produce $l \simeq W$. In reality
the edge roughness coexists with other types of disorder. Reference \cite{Chand} did not specify disorder in measured samples, but Webb mentioned in \cite{Kircze} that the grains
in the rings of work \cite{Chand} were much larger than $W$ (say in Ref. \cite{WebbWashburn} $d_G \simeq 8W$). The grains with $d_G \gg W$ are known as bamboo-like grains \cite{Graham,Austin,Pumarol}. Of course, $d_G \gg W$ and $l \simeq W$ \cite{Chand} means $l \ll d_G$, which suggests that the grain boundaries were not the main source of scattering in work \cite{Chand}.
If the random grain boundaries (or impurities) were the main source of scattering, the
measured persistent current \cite{Chand} would be $\sim (l/L)I_0$ rather than $\sim I_0$ (c.f. Fig. \ref{Fig-10} and Ref. \cite{Feilhauer2}).
What remains is the edge roughness and it indeed explains the mysterious coexistence of results $I_{typ} \simeq I_0 $, $L/l \gg 1$, and $\langle \rho \rangle \propto L$.
\emph{What happens if one adds the bamboo-like grains?}

\begin{figure}[t!]
\centerline{\includegraphics[clip,width=0.85\columnwidth]{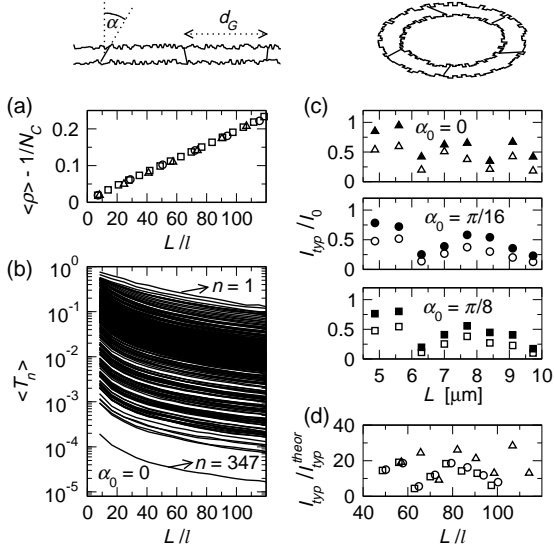}} \caption{
  Transport in Au wires and Au rings with rough edges and bamboo-like grains. The angle $\alpha$ of the grain boundary is chosen at random from the interval $(-\alpha_0,\alpha_0)$, where $\alpha_0$ is the parameter: $\alpha_0 = 0$ means the ideal bamboo shape with the boundary perpendicular to the wire \cite{Graham,Austin,Pumarol}.
  The table shows all parameters and the resulting $l$ and $\xi$. Figure (a) shows the
   mean resistance $\langle \rho \rangle$ versus $L/l$, figure (b) show the transmission $\langle T_n \rangle$ versus $L/l$ for $\alpha_0 = 0$. Open symbols in figure (c) show the typical current $I_{typ}/I_0$ versus $L$ for various $\alpha_0$, the full symbols show the maximum currents.
   Figure (d) shows the $I_{typ}$ data from figure (c) normalized by $I^{theor}_{typ} = 1.6 (e v_F/L) (l/L)$ and plotted in dependence on $L/l$.} \label{Fig-13}
\vspace{-0.5cm}
\end{figure}

Since $d_G \gg W$, we fit $R_G$ to obtain $l \simeq W$. Figure \ref{Fig-13} shows such a study for the same $W$ and similar $L$ as in Ref. \cite{Chand}.
In figure \ref{Fig-13}(a) we see again the diffusive law $\langle \rho \rangle \propto L/l$, but now $l \simeq W$, like in Ref. \cite{Chand}. Figure \ref{Fig-13}(b) shows that the transmission through channels $1$, $2$, and a few more is still large
(between $1$ and $0.1$), though not as large as in the wire with rough edges only [c.f. the right panel of Fig. \ref{Fig-2}(b)]. A suppression of the transmission, caused by a combined effect of the rough edges and bamboo-like grains, is visible for all $347$ channels. Consequently, $l \simeq W$.
 Similarly, the typical currents in figures \ref{Fig-13}(c) and \ref{Fig-13}(d) are suppressed in comparison with the pure edge-roughness case [Fig. \ref{Fig-10}(b)], but they still grossly exceed the value $I^{theor}_{typ} = 1.6 (e v_F/L) (l/L)$.
Figure \ref{Fig-13}(c) presents the maximum currents, because \emph{Ref. \cite{Chand} in fact reported the current amplitudes rather than $I_{typ}$. These amplitudes were between $\sim 0.1I_0$ and $\sim I_0$ and essentially the same show our data} (the full symbols).

\section{V. Summary and concluding remarks}

\subsection{A. Summary}

In our paper, persistent currents in mesoscopic normal-metal rings with disorder due to the rough edges and random grain boundaries have been calculated by means of the single-particle scattering-matrix method. In addition, the diffusive resistance of corresponding metallic wires has been obtained from the Landauer formula and the diffusive electron mean free path has been determined. Our calculations capture two crucial points. First, disorder is described microscopically; we do not rely on the approximation of the spatially homogeneous white noise. Second, our description of the single-electron states in the ring captures the key effect of the Hartree-Fock interaction, the cancelation of the centrifugal force by an opposite oriented Hartree-Fock field.

Our main results (Fig. \ref{Fig-10}) are the following.
If disorder is due to the random grain boundaries, our results for the typical persistent current agree with the white-noise-related formula $I_{typ}\simeq (e v_F/L)(l/L)$ and recent experiments \cite{Bluhm,Bles}. However, if the disorder is due to the rough edges, we find the ballistic-like current $I_{typ}\simeq e v_F/L$ albeit the resistance is diffusive ($\propto L/l$) and $L \gg l$.
In other words, the multichannel disordered metal ring of length $L \gg l$
supports the current $I_{typ} \simeq e v_F/L$, expected to exist only in a single-channel disorder-free ring.
 This finding agrees with experiment \cite{Chand}.

 Thus, figure  \ref{Fig-10} naturally explains the difference between the experiment \cite{Chand} and experiments \cite{Bluhm,Bles}.
 It simply suggests that
disorder in samples of works \cite{Bluhm,Bles} was white-noise-like (most likely mainly due to the random grain boundaries), while disorder in samples
of work \cite{Chand} was likely mainly due to the rough edges. Ideally, the ballistic persistent current is inherent to metallic rings
with rough edges. However, according to our data in figure \ref{Fig-13} , it survives (slightly suppressed) also when the bamboo-like polycrystalline grains are added
in order to emulate the polycrystallinity of the real rings \cite{Chand}.

 The microscopic origin of the ballistic persistent current in metallic rings with rough edges has been explained.
 The ballistic current is mainly due to the electrons that occupy
channel $n=1$. Classically speaking, these electrons move (almost) in parallel with the ring edges and therefore avoid the edge roughness scattering.
 The reason why they move in parallel with the ring edges in spite of the ring geometry is the Hartree-Fock interaction; it acts against the centrifugal force and eliminates the effect of the ring geometry. In terms of classical paths, the electron paths in presence of the Hartree-Fock field are not the straight lines; the field deflects them from the outer ring edge and the resulting electron wave function is centered between the edges almost symmetrically.

Finally, we recall that all our results are universal. The transport results obtained for the grain boundary model
in figure  \ref{Fig-1}(a) hold for any other grain boundary model in which the orientation and positions of the boundaries are random. The transport results obtained for the step-shaped-roughness model in figure \ref{Fig-1}(b) hold also for models with a smoothly varying roughness (section II.C). The universality exists also within our specific roughness model; all our results are robust against the change of parameters  $\delta$, $\Delta x$, $N_c$, $l$, and $L$, if they are plotted in dependence on $L/\xi$. Therefore, a missing information on the nature of disorder in measured samples \cite{Chand,Bluhm,Bles} is not crucial for our conclusions. Anyway, our values of $\delta$ and $\Delta x$ are close to the real ones \cite{Bryan}. In principle, we could attempt to reproduce the measured values of $I_{typ}$ and $l$
exactly by fitting the parameters of disorder. This should make sense if new experiments determine
$I_{typ}$ and $l$ together with the parameters of disorder in measured samples.

\subsection{B. Remark on robustness of the 2D results against 3D effects}

Our results were obtained for the 2D model of figure \ref{Fig-1}, while the experimental samples \cite{Chand,Bluhm,Bles}
 are three-dimensional. We want to point out that the extension of our 2D study to 3D would not change our results remarkably. The effect of 3D can be estimated without an explicit calculation.

In our 2D wire (Fig. \ref{Fig-1}.b) the roughness scattering is due to the wire edges. In real 3D wires  the roughness scattering is in general due to the wire edges (side walls) as well as due to the top and bottom surfaces. In spite of this difference the 3D sample preserves the key feature of our 2D model. Namely, the electrons in the ground 1D channel (now the channel with quantum numbers $n_y = 1$ and $n_z = 1$, where $z$ is the vertical direction) still move almost in parallel with the sample edges and sample surfaces, and therefore avoid the roughness scattering. Thus, the transmission through the ground 1D channel has to be ballistic, similarly as we have seen for the 2D wire (Fig.2). Consequently, the 3D rings have to carry for $L/l >> 1$ the ballistic current $I_{typ} \simeq ev_F/L$, similarly as the 2D rings in figure \ref{Fig-10}(b).

Further, the roughness scattering in 3D does not modifies the mean free path $l$ remarkably in comparison with 2D.  Indeed, in real 3D wires the roughness amplitude (RMS) of the top and bottom surfaces is usually of the order of one lattice constant ($\sim 0.5$nm; see e.g. the paper \cite{Munoz}),  which is far less than the roughness amplitude at the edges (RMS $\sim 5$nm - $10$nm; see the experiment \cite{Bryan} and our present paper). Since the roughness-limited mean free path is proportional to the square of the RMS \cite{Feilhauer1}, the effect of the top and bottom surfaces on the mean free path has to be two orders of magnitude weaker than the effect of the edges. It is thus very likely that the roughness scattering in the 3D wires of reference \cite{Chand} is mainly due to the wire edges.

Finally, unlike the 2D wire in figure \ref{Fig-1}, the edges of the 3D wire are the side walls and the edge roughness at such side walls in general scatters the electrons also in the vertical ($z$) direction. In comparison with our purely 2D scattering, this may decrease the roughness-limited mean free path say by a few tens of percent. However, this cannot affect the ballistic-like motion in the ground 1D channel, responsible for the ballistic current $I_{typ} \simeq ev_F/L$ at $L/l >> 1$.

\subsection{C. Remark on an angle dependence of roughness scattering}

     In the non-interacting-electron model of the 2D ring the wave functions are dominated by the straight-line electron paths \cite{Jalabert,Samokhin}. To incorporate the edge roughness scattering in that model, it was assumed \cite{Samokhin} that any straight-line path which hits the ring edge is reflected diffusively no matter what is the incidence angle (the angle between the path and the edge).  However, a realistic probability of diffusive reflection, derived by Ziman and Soffer \cite{Soffer,Ziman} for
a free wave impinging the surface with uncorrelated roughness, strongly depends on the incidence angle. It is equal to unity
for perpendicular incidence but approaches zero for small incidence angles. Note that
the realistic angle dependence of the edge roughness scattering is inherent to our scattering matrix method.

Indeed, the tendency to a specular reflection at small angles is manifested by the channel transmission $T_n$. Let us look at the right panel of figure 2b in detail.
Classically, the channel number $n$ corresponds to the angle between the classical trajectory and edge, and $n=1$ corresponds to the smallest nonzero
classical angle allowed by the quantum confinement. Consider say $L \simeq 0.25 \xi \simeq 120 l$. In case of the diffusive reflection assumed by work \cite{Samokhin}, for $L/l = 120$ we should observe $T_n \sim l/L \sim 1/120$ for all $n$ including $n=1$. However, this is not the case; the right panel of figure 2b shows that $T_n$ is between 1 and 0.1 for $n =1, 2, \dots, 6$. Evidently, the electron motion in these channels is much more ballistic than diffusive. When this realistic angle dependence is combined with the Hartree-Fock interaction, the metallic rings with rough edges support for $L \gg l$ ballistic current $I_{typ} \simeq e v_F/L$.

 \subsection{D. $T=1$ as a general feature of any diffusive wire and $T_{n=1} \simeq 1$ in the wire with rough edges: Two different things}

     We note that the transmission $T_{n=1} \simeq 1$ in the wire with rough edges (right panel of figure \ref{Fig-2}(b)) has nothing in common with the well-known bimodal distribution $1/\sqrt{(1-T)T^2}$, which exist in any diffusive conductor \cite{vanRossum} and diverges for $T=1$. Transmissions $T$ in the bimodal distribution are the eigen-values of the $t^{+}t$ matrix \cite{vanRossum},
our $T_n=\sum_{m=1}^{N_C}|t_{n,m}|^2$ are the diagonal elements of the $t^{+}t$ matrix.
 In other words, the channels corresponding to the eigen-values $T$ in the distribution $1/\sqrt{(1-T)T^2}$ are the eigen-states of the $t^{+}t$ matrix and the channels corresponding to our diagonal elements $T_n$ are the plane-wave states. This difference deserves a few remarks.

The bimodal distribution $1/\sqrt{(1-T)T^2}$ as a general property of any diffusive conductor with white-noise-like disorder \cite{vanRossum} coexists with the diffusive persistent current $I_{typ} \simeq (ev_F/L)(l/L)$ in the corresponding disordered ring \cite{Cheung}. This means that the eigenvalues $T=1$ in the bimodal distribution do not cause any ballistic persistent current. The reason why the current is diffusive in spite of $T=1$, is most likely that the eigenvalue $T=1$ does not necessarily mean the ballistic transmission (a well known example is the perfect transmission in case of resonant tunneling).

For disorder due to rough edges the situation is fundamentally different. In this case the eigen-values $T$ still follow the bimodal distribution $1/\sqrt{(1-T)T^2}$, however, this has nothing in common with the ballistic-like persistent current found by us. The ballistic-like current is due to the appearance of the diagonal element $T_{n=1} \simeq 1$. Specifically, any wire in the statistical ensemble of wires with rough edges exhibits the diagonal element $T_{n=1} \simeq 1$ independently on the choice of the Fermi energy and wire length. It is easy to check for any of our simulated wires, that the electron plane wave which enters the wire in channel $n=1$ remains (almost) unscattered between any two successive scatterers inside the disordered region. As a result, the ring made of such wire supports the persistent current dominated by the ballistic channel $n = 1$, that is, $I_{typ} \simeq ev_F/L$. In summary, the reason for appearance of $I_{typ} \simeq ev_F/L$ is the ballistic behavior of the diagonal
element $T_{n=1}$; the fact that the bimodal distribution shows eigenvalues $T=1$ is irrelevant.

\subsection{Acknowledgement}

We thank the Texas Advanced Computing Center (TACC) at
the University of Texas at Austin for providing grid resources. We thank for grant VEGA 2/0206/11.

%

\end{document}